\documentclass[draft]{aipproc} 
\layoutstyle{6x9}
\usepackage{amssymb}

\begin{document}

\title[Exploring Quantum Gravity with VHE Instruments]{Exploring Quantum Gravity with Very-High-Energy Gamma-Ray Instruments -- Prospects and Limitations\thanks{This artice has been submitted to AIP Conference Series. After it is published, it will be found at {\tt http://link.aip.org/link/?apc}}
}
% 98.54.Cm Active and peculiar galaxies and related systems (including BL Lacertae objects, blazars, Seyfert galaxies, Markarian galaxies, and active galactic nuclei)
% 04.60.-m Quantum gravity (see also 11.25.-w Strings and branes)
% 04.60.Bc Phenomenology of quantum gravity
% 98.70.Rz gamma-ray sources; gamma-ray bursts
\author{Robert Wagner}{
 address={Max-Planck-Institut f\"ur Physik, D-80805 M\"unchen, Germany},
 %email={robert.wagner@mpp.mpg.de}
}

\begin{abstract}
Some models for quantum gravity (QG) violate Lorentz invariance and predict an
energy dependence of the speed of light, leading to a dispersion of high-energy
gamma-ray signals that travel over cosmological distances. Limits on the
dispersion from short-duration substructures observed in gamma-rays emitted by
gamma-ray bursts (GRBs) at cosmological distances have provided interesting
bounds on Lorentz invariance violation (LIV). Recent observations of
unprecedentedly fast flares in the very-high energy gamma-ray emission of the
active galactic nuclei (AGNs) Mkn 501 in 2005 and PKS 2155--304 in 2006
resulted in the most constraining limits on LIV from light-travel observations,
approaching the Planck mass scale, at which QG effects are assumed to become
important. I review the current status of LIV searches using GRBs and AGN flare
events, and discuss limitations of light-travel time analyses and prospects for
future instruments in the gamma-ray domain.
\end{abstract}

\maketitle

\section{Introduction}
Currently we have two extremely successful fundamental theories at hand in
physics: general relativity, describing space, time and gravitation; and
quantum theory, describing electroweak and strong interactions by means of
quantum field theories.  The interface of both theories was not explored for a
long time, because there have been no experiments that could probe regimes in
which both quantum and gravitational effects become important. Still, not only
from a philosophical point of view, physicists aspire a unified theory of both,
which is usually called quantum gravity (QG). In standard relativistic quantum
field theory (QFT), space-time is considered a fixed arena in which physical
processes take place. Quite in contrast, QG theories require drastic
modifications of space-time when structures approach the QG mass scale
\cite{ame98,gar98,gam99}, often identified with the Planck mass
$M_\mathrm{Planck} \equiv \sqrt{\hbar c/G_N} = 1.22\times 10^{19}\,
\mathrm{GeV}$. However, the energies at which QG effects lead to deviations
from conventional quantum mechanics might lie well below the QG energy scale.
As a non-renormalizable interaction gravity may leave distinctive imprints also
at energies much below the preferred QG scale, if violating any fundamental
symmetry. Thus, we might spot QG effects not only on extreme energy scales, but
use GeV/TeV photons as
``low-energy'' probes.

An energy dependence of the speed of light in vacuum may arise from photon
propagation through a gravitational medium containing quantum fluctuations on
distance scales on the order of the Planck length $l_\mathrm{Planck} \equiv
\sqrt{\hbar G_N/c^3}\approx 10^{-33}\,\mathrm{cm}$ and on timescales $\approx
M_\mathrm{Planck}^{-1}$. While being smooth at large distances, space-time at
short distances of the order of $l_\mathrm{Planck}$ might show a complex,
foamy, structure due to quantum fluctuations \cite{whe63,haw82,ell84}. Such
Planck-size topological defects, viz. black holes with microscopic event
horizons, appearing and evaporating spontaneously, might lead to quantum
decoherence and Lorentz invariance (LI) violation.
There are numerous formal mathematical approaches to QG. Among the most popular
ones are loop QG \cite{gar94,gam99}, which is the canonical non-perturbative
quantization of general relativity, and superstring theory, which originally
has been developed to describe the behavior of hadrons. (See \cite{smo03} for a
review on QG theories). In Liouville string theory \cite{ell99}, modified
dispersion relations (MDR) are enabled by higher-order derivatives in effective
Maxwell and Dirac equations.

It has been pointed out by Amelino-Camelia et~al.\,\cite{ame98} that different
approaches to QG lead to a similar description of first-order effects of such a
time dispersion:
\begin{equation}
\Delta t \simeq \xi \frac{E}{E_\mathrm{QG}} \frac{L}{c}
\end{equation}
where $\Delta t$ is the time delay relative to the standard speed of light,
$\xi$ is a model-dependent factor of the order 1, $E$ is the energy of the
observed radiation, and $L$ is the distance traversed by the radiation. Thus,
LI violation (LIV) is expected as a generic signature of approaches to QG. A
rather general review on the status of LIV tests is given by
Mattingly~\cite{mat05}.

A number of predicted new phenomena lead to testable astrophysical effects.
Dispersion \cite{ame98} and vacuum birefringence \cite{gam99} (leading to
polarization-dependent LIV) are purely kinematical effects and require only an
MDR and a standard definition of the group velocity. Below, we will
particularly elaborate on astrophysical time-of-flight tests of the dispersion
relation of photons. Other effects, like anomalous threshold reactions,
threshold shifts in standard reactions, or reactions affected by ``speed
limits'' (e.g. synchrotron radiation), need additional assumptions on
energy/momentum conservation and dynamics, or on effective QFT, and will not be
discussed here. Note that a non-constant speed of light may also occur with
maintained Lorentz invariance, e.g., in doubly-special relativity \cite{mag02};
this underlines the necessity for astrophysical tests of LIV (LIV tests on
Earth would yield negative results!).

In the remaining sections, we will review astrophysical probes of QG with high
energy $\gamma$-rays.

\section{Probing Quantum-Gravity with Photon Time-of-Flight measurements}
\begin{figure}
\resizebox{.5\textwidth}{!}{\includegraphics{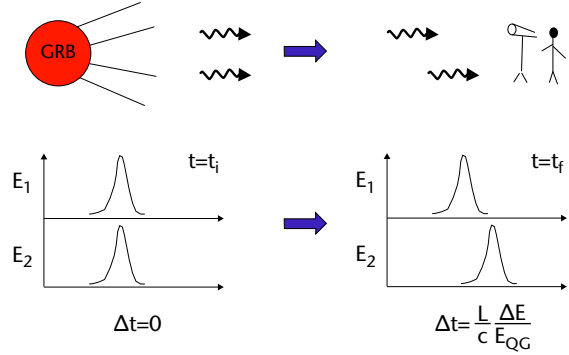}}
\caption{\label{fig:morselli} In photon time-of-flight
measurements, it is assumed that the energy dispersion at the
source is known or negligible. Any additional dispersion can then
be attributed to the dispersive properties of the vacuum, and this
dispersion can be measured by the observer. Original drawing
courtesy A.~Morselli.}
\end{figure}

When looking for QG phenomena, we expect deviations from QFT, presumably
suppressed by some power of the Planck mass (or the QG mass scale). From a
purely phenomenological point of view \citep{ame98}, such effects can be
treated using a perturbative expansion, assuming the involved energies $E \ll
M_\mathrm{Planck}$:
\begin{equation}
c^2 p^2 = E^2 \left( 1+ \xi
\left(\frac{E}{M_\mathrm{Planck}}\right) + \zeta
\left(\frac{E^2}{M^2_\mathrm{Planck}}\right) + \mathcal{O}
\left(\frac{E^3}{M^3_\mathrm{Planck}}\right) + ...\right)
\label{eq:2}
\end{equation}
Note that an explicit breaking of LI is expected at the Planck mass scale. A
linearly-deformed dispersion relation arises, e.g., in noncritical Liouville
string theory, while in critical string theory, only quadratic deviations are
expected \cite{ell08}.  An MDR implies an energy-dependent speed of light,
\begin{equation}
v = \frac{\partial E}{\partial p} \simeq c \left( 1- \xi
\left(\frac{E}{M_\mathrm{Planck}}\right)\right)
\end{equation}
or, in other words, the vacuum acquires non-trivial optical properties, namely
a refractive index $v(E)=c/n(E)$. This translates into a delay, including
cosmological effects,
\begin{equation}
\Delta t = \xi \frac{\Delta E}{M_\mathrm{QG}} \frac{L}{c} = \xi
\frac{\Delta E}{M_\mathrm{QG}} H_0^{-1} \int \mathrm{d}z/h(z).
\end{equation}
To measure any such time delay, very fast transient astrophysical phenomena are
required, which provide a time stamp for the simultaneous emission of photons
of different energies (Fig.~\ref{fig:morselli}). The corresponding figure of
merit for QG tests is given by the sensitivity to a given QG mass scale,
\begin{equation}
M_\mathrm{QG} = \xi  \frac{L}{c}  \frac{E}{\Delta t}
\end{equation}
The photon energy $E$ acts as lever arm, implying that both an instrument for
measuring $\gamma$-ray energies as high as possible, but also astrophysical
sources that deliver these energies are advantageous. But also the time
resolution of the instrument and the fineness of the structures in the
$\gamma$-ray signal, described by $\Delta t$, are decisive, and the source has
to provide enough photon statistics per time as to allow for a binning as small
as possible. $L$ denotes the luminosity distance of the source.

\section{Quantification of Time Delays}
In the X-ray to VHE $\gamma$-ray domain, photon data are characterized by
arrival directions, energies, and arrival times of X-rays, gamma rays, or
gamma-ray candidates. A very basic method of assessing any time delay of very
high energy photons, motivated also by the sparse statistics at these energies,
relies on finding peaks in the cross-correlation function of light curves of
different photon energy bandpasses \cite{ban97,bil99,mag07}. Note, however,
that such analyses do not take into account the full available information each
individual photon carries \cite{mag08}. Therefore, more appropriate methods
have been developed, as summarized in the following.
\begin{itemize}
\item In the energy-cost-function method \cite{mag08}, it is
assumed that traveling through any dispersive medium will increase
the spread of the photon arrival time distribution.\footnote{Note
that the dispersion needs not be source-extrinsic, thus even
dispersive regions inside a $\gamma$-ray source can be modeled
using the ECF approach.} Thus, by minimizing the spread of an
observed, dispersive-broadened pulse by an ``energy cost
function'', the original pulse and the dispersion can be
reconstructed \cite{mag08,sca06}.
\item A log-likelihood
fit, describing the flare peak form $\sigma_E$ (which might be
assumed, e.g., as Gaussian pulse), the instrumental energy
smearing $G$, the energy spectrum $\Gamma(E_s)$ of the flare, and
the observed photon arrival times $F_s(t_s)$ can be applied to the
unbinned data \cite{lam07,mar08}, e.g.,
\begin{equation}
\frac{{\mathrm d}P}{{\mathrm d}E {\mathrm d}t} = k \int_0^\infty
\Gamma(E_s) G(E-E_s, \sigma_E(E_s)) F_s(t_s) {\mathrm d}E_s.
\end{equation}
\item Cross-correlation of oversampled light curves \cite{lix04}.
This ``modified cross-correlation function'' (MCCF) method allows
for resolving features (as time delays) in the light curves well
below the duration of the flux bins.
\item A wavelet analysis suppresses noise in the data and
identifies ``genuine variation points'' \cite{ell03}, which can
be attributed to flares and can be used as time stamps in
different energy bandpasses. This method is particularly useful if
the observational data are noisy and the (multiple or complex)
flare structure is not very significant.
\end{itemize}

\section{Present Reach using High-Energy Gamma-Rays}
Pulsars have very distinct periodic light-curve forms, and due to the
periodicity, these light curves can be sampled very accurately. In observations
of photons up to $E=2\,\mathrm{GeV}$ from the Crab pulsar, at a distance of
$2.0\pm0.5$~kpc \cite{kap08}, using the sub-millisecond timing of EGRET,
$\Delta t=10^{-3}\,\mathrm{s}$ was achieved. With these numbers, a sensitivity
to $M_\mathrm{QG} \approx 10^{15}\,\mathrm{GeV}$ can be achieved. For active
galactic nuclei (AGN; typical distances: in excess of 100 Mpc $\cong$
$10^{27}\mathrm{cm}$), Imaging Air Cerenkov Telescopes (IACT
\cite{jritm,edotm,mbetm}) currently may observe bright $\gamma$-ray flares with
energies up to around $E_\gamma \approx 10\,\mathrm{TeV}$ with sufficient
statistics and a time resolution (given by the AGN flare timescales) of
$\approx 50-100\,\mathrm{s}$. This enables a sensitivity up to the Planck
energy scale, $M_\mathrm{QG} \approx 10^{19}\,\mathrm{GeV}$. In the absence of
a linear term in the MDR, the quadratic reach would be around
$M_\mathrm{QG}=6\times10^9\,\mathrm{GeV}$. Gamma-ray bursts, at distances
beyond 7000 Mpc, will be routinely observed by {\it Fermi} \cite{jmetm} at
energies up to $\mathcal{O}(\mathrm{GeV})$, which will yield a sensitivity of
$M_\mathrm{QG} \approx 10^{19}\,\mathrm{GeV}$.  Note, however, that current GRB
studies are restricted to $M_\mathrm{QG}=10^{17}\,\mathrm{GeV}$, as until
recently only $\Delta E\approx 100$\,keV was available
($M_\mathrm{QG}=10^{6}\,\mathrm{GeV}$ in the quadratic case).

\section{Astrophysical VHE Gamma-Ray Sources}
\vspace*{-.2cm}
\subsection{Pulsed emission from the Crab nebula} 
\vspace*{-.2cm}
The Crab pulsar has a pulsation period of 33.18 ms. As its pulsations are well
aligned in time from radio through X-ray wavelengths, it seems likely that
photons of different energies are produced nearly simultaneously. Using the
sub-ms timing of the EGRET experiment, it could be shown \cite{kaa99} that $>2$
GeV photons trail those at $70-100$ MeV by no more than 0.35~ms (95\% c.l.),
implying $M_\mathrm{QG} > 1.8 \times 10^{15}\,\mathrm{GeV}$.  Recently, pulsed
emission from the Crab was also detected at energies up to 60 GeV
\cite{magcr,shatm}: About 8,500 $\gamma$-ray events form a signal at the
6.4-$\sigma$ significance level. This observation might yield a lower limit for
quantum gravity effects of $M_\mathrm{QG} \approx 1.2 \times
10^{16}\,\mathrm{GeV}$ \cite{magc2}, exceeding the limit given by the
EGRET-Crab analysis.

\vspace*{-.2cm}
\subsection{Gamma-Ray Bursts}
\vspace*{-.2cm}
Gamma-ray bursts (GRBs; \cite{pir03}) have rather complex individual time
profiles. Once a relativistic fireball is created in the gigantic explosions
thought to be the origin of GRBs, the physics ought to be insensitive to the
nature of the progenitor object. GRBs have been found at redshifts exceeding
$z=5$ (7\% of the presently known GRBs; the average redshift is $z=2.8$).
Thus GRBs seem to be a natural proxy to test QG predictions. The key issue is
to distinguish possible QG signatures from intrinsic delays, particularly as
delays in GRBs have been observed. QG effects should increase predictably with
the distance of the individual GRB.
% (in the absence of any cosmological evolution effect). 
%
The accessible energies, however, have been smaller than or at most some GeV;
IACT, which are sensitive in the above 30~GeV region, are still awaiting the
first GRB detection \cite{maggr,hesgr}. An additional complication arises from
the fact that establishing the distance of GRBs is not trivial and works only
for a fraction of the observed GRBs. The strongest QG constraint from an
individual GRB is that from GRB 930229 \cite{sch99}, where during a rise time
of $220\pm30 \mathrm{\mu s}$ the dispersion of photons was measured to be
$\Delta t < 25$\,ms at energies $\Delta E=30 \mathrm{keV} - 80 \mathrm{MeV}$,
yielding $M_\mathrm{QG}>8.3 \times 10^{16}\,\mathrm{GeV}$.  The caveat of this
observation is that GRB 930229 has no measured redshift(!), thus with an
assumed peak flux of $10^{57} $ photons s$^{-1}$ and the peak width--redshift
relation \cite{fen92} a distance $D=260$\,Mpc was inferred. For GRB 051221A,
which occurred in a galaxy at $z=0.55$, no time delay of photons between 15 and
350 keV was observed \cite{mar06}, leading to a limit of $M_\mathrm{QG}>6.6
\times 10^{16}\,\mathrm{GeV}$.  

A more robust method is to study large samples of GRB. In \cite{ell06}, a total
of 36 GRBs observed by BATSE (9 bursts), {\it HETE-2} (15 bursts), and {\it Swift} (11
bursts) up to $z=6.3$ was used.  The energy difference of photons in that
sample is around 200 keV, and to account for possible source effects, a
``universal source uncertainty'' of 54 ms was accounted for. A limit of
$M_\mathrm{QG}>0.9 \times 10^{16}\,\mathrm{GeV}$ was deduced.

\paragraph{{\it Fermi} Large-Area Tracker (LAT) and Gamma-Ray Bursts}
In LAT, around 250 bursts per year are expected \cite{omotm} to be detected,
and about 50 of them will survive a high-energy cut, i.e., contain $>550$~MeV
photons.  LAT itself should detect around 15 bursts without Gamma-ray Burst
Monitor triggers. Kuehn et~al.~\cite{kue07} simulated the LAT response to GRB
photons, focusing on the prompt GRB emission. Due to the wide field of view,
LAT in contrast to the TeV instruments is quite unique in detecting prompt GRB
emission. For the simulations, spectral indices have been taken from BATSE and
extrapolated to LAT energies. Even with a binned analysis, a simultaneous fit
to multiple GRBs (at $z=1$) was shown to have a sensitivity at $10^{19}$~GeV.
An alternative method \cite{sca06} also has this sensitivity.  Recently, the
LAT collaboration reported the first detection of high-energy photons from
three GRBs \cite{bou08}, and used the strongest burst, GRB 080916C, to derive a
lower QG energy scale limit of $M_\mathrm{QG}>(1.5 \pm 0.2) \times 10^{18}\,
\mathrm{GeV}$.

\subsection{Flares from Active Galactic Nuclei}
The study of high energy ($E\gtrsim 100$ MeV) $\gamma$-ray emission from active
galactic nuclei (AGN) is one of the major goals of space-borne and ground-based
$\gamma$-ray astronomy.

The vast majority of the currently known VHE $\gamma$-ray emitting AGNs
\cite{wag07} are blazars \cite{pad95},\footnote{See {\tt
http://www.mpp.mpg.de/$\sim$rwagner/sources/} for an up-to-date source list.} a
subclass characterized by high variability over the whole electromagnetic
spectrum, a characteristic peak in their spectral energy distributions (SEDs)
from nonthermal synchrotron emission, and a second peak at $\gamma$-ray
energies. In synchrotron-self-Compton (SSC) models it is assumed that the
observed $\gamma$-ray peak is due to the inverse-Compton (IC) emission from the
accelerated electrons up-scattering previously produced synchrotron photons to
high energies \cite{mar92}. In hadronic models, instead, interactions of a
highly relativistic jet outflow with ambient matter, proton-induced cascades,
or synchrotron radiation off protons, are responsible for the high-energy
photons.

The duty cycle of {\it Fermi} enables a quasi-continuous
monitoring of blazar flaring activity (also monitoring programs by
IACTs are in place, \cite{sattm}), while the high sensitivity of
current IACT is advantageous for understanding the particle
acceleration mechanisms in flares, the origin of the high-energy
$\gamma$-rays, as well as the relations between photons of
different energies (from radio to VHE; particularly between the
two emission populations \cite{wag08}).

\vspace*{-0.2cm}
\subsection{The Whipple QG Limit from a Mkn 421 Flare}
The first rapid flare from a TeV blazar was observed by the Whipple telescope
from Markarian (Mkn) 421 ($z=0.030$) on 1996 May 15 \cite{gai96}. The peak of the
flare is concentrated in one 280~second bin, and as there were no 
$>2$~TeV $\gamma$-rays observed outside this bin, it was concluded that the
$\gamma$-rays $<1$~TeV and $>2$~TeV are in step with $>$95\%
probability \cite{bil99}. This observation translates in a lower
limit $M_\mathrm{QG}>4\times10^{16}~\mathrm{GeV}$.

\vspace*{-0.2cm}
\subsection{The July-2005 Flares of Mkn 501}
\begin{figure}
\resizebox{.7\textwidth}{!} {\includegraphics{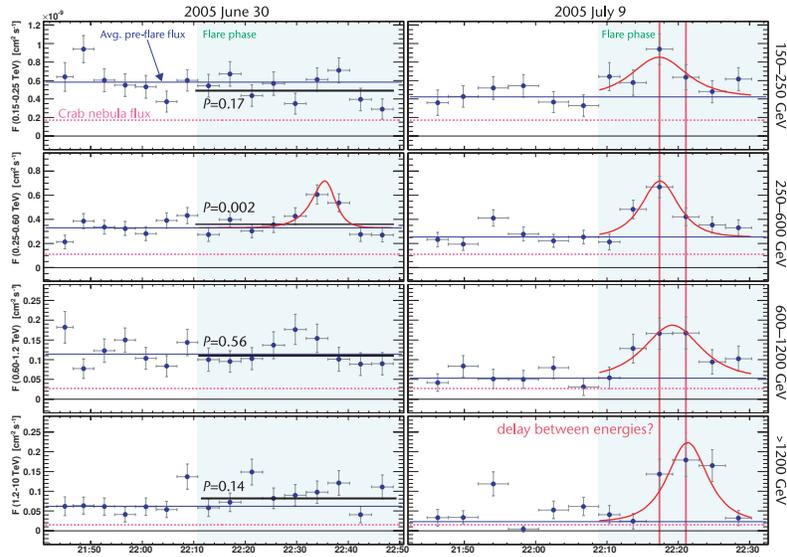}}
\caption{\label{fig:magic} Light curves for the observations on
2005 June 30 (left panels) and July 9 (right panels)
\cite{mag07} with a time binning of 4 minutes, and separated in
different energy bands ranging from 150 GeV to $>$1.2 TeV. For
comparison, the Crab nebula emission level is shown. The data are
divided in into `stable' and `variable' emission. The
probabilities given in the left panels are for fits with constant
lines in the `variable' region. Only the constant-line fit in the
energy range 0.25--0.6 TeV is not satisfactory. In the right
panels, the `variable' regions were fitted with a flare model. A
more quantitative photon-by-photon analysis of these data was
performed in \cite{mag08}.}
\end{figure}

MAGIC \cite{jritm} recorded fluxes exceeding four times the Crab-nebula
flux from Mkn 501 ($z=0.034$) in 2005, and revealed rapid flux changes with
doubling times as short as 3 minutes or less \cite{mag07}. For the first time,
short ($\approx$~20 minute) VHE $\gamma$-ray flares with a resolved time
structure could be used for detailed studies of particle acceleration and
cooling timescales. The two flares behaved differently (Fig.~\ref{fig:magic}):
While the 2005 June 30 flare is only visible in 250\,GeV--1.2\,TeV, the 2005
July 9 flare is apparent in all energy bands (120~GeV to beyond 1.2~TeV).
Additionally, at a zero-delay probability of $P=0.026$, a marginal time delay
$\Delta t=\tau_l E$ with $\tau_l = (0.030\pm0.012) \mathrm{s\,GeV}^{-1}$
towards higher energies was found using two independent analyses, both
exploiting the full statistical power
of the dataset (see \cite{mag08,mar08} for details). For a
quadratic effect (vanishing first-order term in eq.~\ref{eq:2}),
$\Delta t=\tau_q E^2$ with $\tau_q = (3.71\pm2.57)\times
10^{-6}\,\mathrm{s\,GeV}^{-2}$ is obtained correspondingly.
Several explanations for this delay have been considered up to
now:
\begin{enumerate}
\item Electrons inside the emission region moving with constant
Doppler factor are gradually accelerated to energies that
enable them to produce corresponding $\gamma$ rays \cite{mag07}.
\item The $\gamma$-ray emission has been captured in the initial
acceleration phase of the relativistic blob in the jet, which at
any point in time radiates up to highest $\gamma$-ray energies
possible \cite{bed08}.
\item An one-zone SSC model, which invokes a brief episode of
increased particle injection at low energies \cite{mas08}.
\item When assuming a simultaneous emission of the $\gamma$-rays
(of different energies) at the source at $z=0.034$, an observed
time delay can be converted in QG mass scales $M_\mathrm{QG1} =
1.445\times10^{16}\mathrm{s}/\tau_l$ and $M_\mathrm{QG2} =
1.222\times10^{8}(\mathrm{s}/\tau_q)^{1/2}$, respectively. As the
measured time delay is marginal, they translate in a lower limit
of $M_{\mathrm{QG1}} > 0.21 \times 10^{18}$~GeV (95\% c.l.) for
the linear case and $M_{\mathrm{QG2}} > 0.26 \times 10^{11}$~GeV
(95\% c.l.) for a quadratic dependence on energy \cite{mag08}.
Results for the 2005 June 30 exhibit a similar sensitivity, but
this flare is not very significant. Both limits increase further
if any delay towards higher energies in the source itself is
present.
\end{enumerate}

\subsection{The 2006 Giant PKS 2155--304 Flare}
In the night of 2006 July 28, the H.E.S.S. array \cite{edotm}
observed a giant ($>10$ Crab) outburst of the blazar PKS
2155--304, almost four times as distant as Mkn 421 and Mkn 501
($z=0.116$). Also in this case, fast variability on the order of 3
minutes and an energy coverage up to a few TeV is given
\cite{hes08}. For searching for time delays, a MCCF method
\cite{lix04} and a wavelet method were employed, both yielding a
zero time delay within errors, and thus $\tau_l >
0.073\,\mathrm{s\,GeV}^{-1}$ and $\tau_q = 45 \times
10^{-6}\,\mathrm{s\,GeV}^{-2}$, respectively (95\% c.l.). This
translates into lower limits of $M_{\mathrm{QG1}} > 0.52 \times
10^{18}$~GeV and $M_{\mathrm{QG2}}
> 0.014 \times 10^{10}$~GeV.

\begin{figure}
 \resizebox{.7\textwidth}{!}
  {\includegraphics{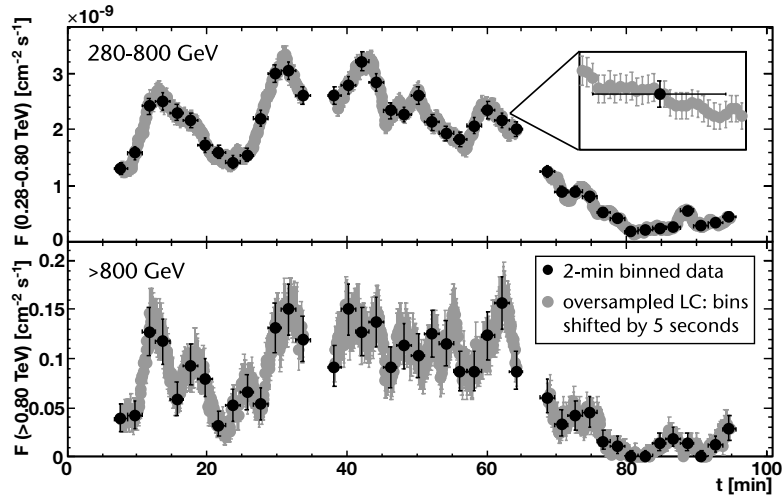}}
\caption{\label{fig:hess} The PKS~2155--304 flare \cite{hes08}.
Black points show the light curves measured between $280-800$ GeV
(upper panel) and $>$800 GeV (lower panel), binned in two-minute
time intervals. Gray points show the oversampled light curve using
the MCCF method \cite{lix04}, for which the two-minute bins are
shifted in units of five seconds. The inlay in the upper panel
illustrates this in a zoom, where the horizontal error bar shows
the duration of the bin in the original light curve.}
\end{figure}

\subsection{Source-Intrinsic Lags and the Blazar Ensemble 2008}
Most acceleration and cooling mechanisms scale in energy, although
there is no definite need for a strong energy dependence. In X-ray
observations, however, soft and hard lags have been observed (most
recently, e.g., by \cite{fos08}), and lead to characteristic
loops in flux--spectral hardness plots \cite{kir98}. Those
diagnostic plots may also be used in the VHE regime \cite{geo01}.
Intrinsic lags may strongly depend on the individual blazar and
even on the individual flare, as the three AGN outburts used for
QG searches have proven. With our current understanding of blazar
physics, intrinsic delays cannot be corrected for individually.
They will, however, not scale with the source distance, so robust
conclusions on QG effects ultimately require the study of samples
of very fast flaring objects at different redshifts. These are
expected to be observed by the current and next-generation
\cite{testm} IACTs and {\it Fermi} thanks to the high sensitivity of
these instruments. Note that multi-wavelength observations of
flares are complementary to QG searches and essential, as they
will allow to study possible source-intrinsic mechanism and
effects and enhance the understanding of AGN source physics. Any
such understanding will enhance QG limits.

\begin{figure}
\resizebox{.7\textwidth}{!} {\includegraphics{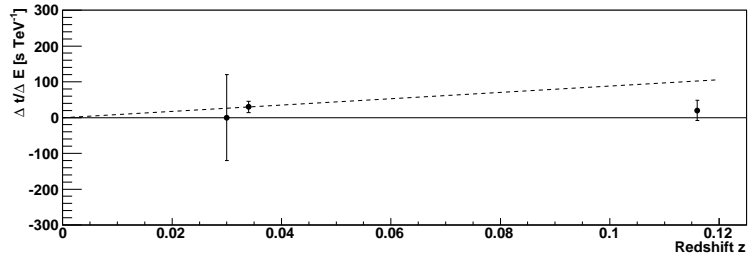}}
\caption{\label{fig:qg} The blazar flare ensemble 2008: Three
limits on the energy dispersion as measured in flares of Mkn 421,
Mkn 501, and PKS 2155--304. If the marginal time delay in Mkn 501
is extrapolated, it is found to be incompatible with the null
result for PKS 2155-304.}
\end{figure}

Note further that obviously source-instrinsic and QG effects may
cancel out, leading to an observation of a null effect. This
scenario has been dubbed ``conspiracy of Nature'' \cite{bil99}
and while it is still conceivable that both effects do exhibit the
same timescale, but the opposite sign, also here a sample of
blazar flares from blazars at different redshifts will help to
separate an eventual QG effect from blazar physics.
Fig.~\ref{fig:qg} shows the current blazar flare ensemble observed
in VHE $\gamma$-rays.

\section{The Future: Prospects and Limitations}
\vspace*{-.3cm}
High energy observations of fast pulses enable measurements of the
energy dependence of the speed of light. Such a possible LIV is
predicted by many QG models. An energy-dependent lag of TeV
$\gamma$-rays on the 2.5-$\sigma$ level has been observed in a
flare of the blazar Mkn 501. The constraints derived from the
$\approx 4$ times more distant blazar PKS 2155--304, however, are
consistent with a null result, as are robust ensemble analyses
from GRBs at a lower $M_\mathrm{QG}$ sensitivity. All in all, VHE
observations of blazars constitute the most stringent limits on a
linear (PKS 2155--304) and quadratic (Mkn 501) energy dependence
in the speed of light. Future improvements in VHE searches may be
expected from:
\begin{itemize}
\item[$\Delta t$ --] Even the high time-resolution IACTs have not
yet explored the shortest blazar variability timescales, although
high photon statistics are available; e.g., more than 10,000
photons in the PKS 2155--304 flare. Thus, only ``brighter flares''
or even more sensitive instruments will help to improve on $\Delta
t$. Observing faster variations, or flares/subflares with shorter
timescales (say 30 seconds instead of currently 3 minutes), might
increase the QG sensitivity by about {\bf $\times 5$}. GRBs with
their fine time structures are interesting to observe,
particularly with {\it Fermi}.
\item[$\Delta E$ --] Gaining access to higher energy $\gamma$-rays
enlarges the lever arm in the QG sensitivity. However, the bright
VHE blazars usually exhibit rather steep spectra, $\sim E^{-2.5}
... E^{-3.5}$ \cite{wag07}, thus it is rather unlikely to gain
an order of magnitude more sensitivity, perhaps a factor {\bf
$\times 3$} is feasible.
\item[$L$ --] It it difficult to improve much using the known
bright, close-by blazars, as we have not yet seen flares exceeding
the 15-Crab level, even not during the giant 1997 flare of Mkn 501
\cite{aha99}; More distant blazars, however, seem to exhibit steep
spectra (e.g. 3C~279 \cite{mag3c,jritm}), which are expected due
to $\gamma$-ray attenuation by the extragalactic background light
\cite{dmatm,hau01,mag3c,hes06}. A major flare in, e.g.,
3C~279, extending the distance from $z=0.12$ to 0.54, 
corresponds to a gain by {\bf $\times 5$} in sensitivity. As
discussed above, for robust ensemble studies, we need to find
flares at various redshifts; thus, improved IACT facilities
\cite{testm} to reach lower energies than the currently reachable
($\approx 30$~GeV by MAGIC; $\approx 100$~GeV by
VERITAS/H.E.S.S.).
\end{itemize}
The overall increase to be hoped for with future IACT observations
is thus about {\bf $\times 70$}.

{\it Fermi} will regularly detect distant ($z>1$) GRBs, thus
collecting statistics and be able to build a robust GRB sample.
This will increase the sensitivity for such a robust GRB study by
about {\bf $\times 20$}.
\vspace*{-.5cm}
\begin{theacknowledgments}
\vspace*{-.3cm}
My thanks go to A. S. Sakharov and the referee 
for helpful comments regarding this manuscript, as well as to the MAGIC group at
MPI for Physics for its excellent support. I am grateful for financial support
by the Max Planck Society.
\end{theacknowledgments}
\vspace*{-.3cm}
\bibliographystyle{aipprocl}

\end{document}